\documentclass[authoryear,a4paper,review]{elsarticle} 
\pdfoutput=1 

\journal{arXiv}

\usepackage{amsmath,amssymb,graphicx,hyperref,setspace} 

\usepackage{natbib}
\usepackage[left=1in,right=1in,top=1.4in,bottom=1.4in]{geometry}
\hypersetup{
	pdftitle={A note on the option price and `Mass at zero in the uncorrelated SABR model and implied volatility asymptotics'},
	pdfauthor={Jaehyuk Choi, Lixin Wu},
	pdfkeywords={Stochastic volatility, SABR model, CEV model, Gauss-Hermite quadrature},
	colorlinks=true,
	linkcolor=red,
	citecolor=blue,
	urlcolor=blue,
	bookmarksnumbered=true,
	pdfstartview=
}

\date{March 26, 2021}

\newcommand{\betac}{\beta_\ast}
\newcommand{\betacpow}[1]{\beta_\ast^{#1}}
\newcommand{\vov}{\nu}

\newcommand{\bs}{\textsc{bs}}
\newcommand{\cev}{\textsc{cev}}
\newcommand{\ncx}{{\chi^2}}
\newcommand{\sabr}{\textsc{sabr}}

\newcommand{\qtext}[2][\quad]{#1\text{#2}#1}

\begin{document}
\begin{frontmatter}
\title{A note on the option price and `Mass at zero in the\\ uncorrelated SABR model and implied volatility asymptotics'}

\author[phbs]{Jaehyuk Choi\corref{corrauthor}\footnote{The source code used in this study can be found at \url{https://github.com/PyFE/PyfengForPapers}}}
\ead{jaehyuk@phbs.pku.edu.cn}

\author[hkust]{Lixin Wu}
\ead{malwu@ust.hk}

\cortext[corrauthor]{Corresponding author \textit{Tel:} +86-755-2603-0568, \textit{Address:} Rm 755, Peking University HSBC Business School, University Town, Nanshan, Shenzhen 518055, China}

\address[phbs]{Peking University HSBC Business School,\\ University Town, Nanshan, Shenzhen 518055, China}
\address[hkust]{Department of Mathematics, The Hong Kong University of Science and Technology,\\ Clear Water Bay, Kowloon, Hong Kong, China}

\begin{abstract}
\citeauthor{gulisashvili2018mass} [Quant. Finance, 2018, 18(10), 1753--1765] provide a small-time asymptotics for the mass at zero under the uncorrelated stochastic-alpha-beta-rho (SABR) model by approximating the integrated variance with a moment-matched lognormal distribution. We improve the accuracy of the numerical integration by using the Gauss--Hermite quadrature. We further obtain the option price by integrating the constant elasticity of variance (CEV) option prices in the same manner without resorting to the small-strike volatility smile asymptotics of \citeauthor{demarco2017shapes} [SIAM J. Financ. Math., 2017, 8(1), 709--737]. For the uncorrelated SABR model, the new option pricing method is accurate and arbitrage-free across all strike prices.
\end{abstract}
\end{frontmatter}

\section{Introduction} \noindent
The stochastic-alpha-beta-rho (SABR) model proposed by \citet{hagan2002sabr} is one of the most popular stochastic volatility models adopted in the financial industry thanks to the availability of an approximate equivalent Black--Scholes (BS) volatility formula (hereafter, the HKLW formula). Despite its enormous successes, the model still poses challenges for enhancements. See \citet{antonov2019modern} for an extensive literature review.

The processes for the price and volatility under the SABR model are respectively given by
\begin{gather} \label{eq:sabr_sde}
dX_t = Y_t\, X_t^\beta\, dW_t \quad (X_0 = x_0) \qtext{and} d Y_t = \vov \, Y_t\, dZ_t \quad (Y_0 = y_0),
\end{gather}
where $\vov$ is the volatility of volatility, $\beta$ is the elasticity parameter, and $W_t$ and $Z_t$ are the (possibly correlated) standard Brownian motions. We will denote the time-to-maturity of the option by $T$ and the strike price by $K$. We also denote $\betac = 1 - \beta$ for simplicity.

The uncorrelated SABR model, where $W_t$ and $Z_t$ are independent, draws attention thanks to the analytical tractability of the constant elasticity of variance (CEV) model,
$$
dX_t = \sigma\, X_t^\beta\, dW_t \quad (X_0 = x_0).
$$
Under the CEV model, the call option price\footnote{The put option price is given by
$$P_\cev(\sigma) = K \, \bar{F}_\ncx \left(\frac{x_0^{2\betac}}{\betacpow{2}\sigma^2 T};\, \frac1{\betac},\frac{K^{2\betac}}{\betacpow{2}\sigma^2 T}\right) - x_0 F_\ncx \left(\frac{K^{2\betac}}{\betacpow{2}\sigma^2 T}; \,2+\frac1{\betac},\frac{x_0^{2\betac}}{\betacpow{2}\sigma^2 T}\right)$$
}
is given by
\begin{gather}
C_\cev(\sigma) = x_0 \bar{F}_\ncx \left(\frac{K^{2\betac}}{\betacpow{2}\sigma^2 T}; \,2+\frac1{\betac},\frac{x_0^{2\betac}}{\betacpow{2}\sigma^2 T}\right) - K \, F_\ncx \left(\frac{x_0^{2\betac}}{\betacpow{2}\sigma^2 T};\, \frac1{\betac},\frac{K^{2\betac}}{\betacpow{2}\sigma^2 T}\right), \label{eq:cev_call}
\end{gather}
where $F_\ncx(\,\cdot\,;r,x')$ and $\bar{F}_\ncx(\,\cdot\,;r,x')$ are respectively the cumulative distribution function (CDF) and complementary CDF of the non-central chi-squared distribution with degrees of freedom $r$ and non-centrality parameter $x'$. The formula is arbitrage-free since they are obtained with the absorbing boundary condition at the origin. 

The uncorrelated SABR model can be interpreted as the CEV model with a stochastic time change~\citep{islah2009sabr-lmm}. The SABR option price, for example, is the expectation of the CEV price over the normalized integrated variance, $V$:
\begin{equation} \label{eq:call_sabr}
C_\sabr (y_0, \vov) = \mathbb{E}\left(C_\cev \left(y_0\sqrt{V}\right)\right), \qtext{where}
V:=\frac{1}{T}\int_0^T e^{2\vov Z_t-\vov^2 t} dt.
\end{equation}
Here, we have omitted the dependency of $C_\sabr$ on the other variables, $x_0$, $\beta$, $K$, and $T$. The probability density of $V$ is available as an integral representation~\citep[Eq.~(4.1)]{matsuyor2005exp1}. The complexity of the density function, however, makes the numerical evaluation of Eq.~\eqref{eq:call_sabr} prohibitive.

\section{Mass at Zero using Gauss--Hermite Quadrature} \noindent
\citet{gulisashvili2018mass} have managed to obtain an approximation to the mass at zero, $m_\sabr = \mathbb{P}(X_0=0)$, in small- and large-time limits. As ``the original motivation of the paper,'' they ultimately use the obtained mass at zero for the arbitrage-free implied volatility formula of \citet{demarco2017shapes} (hereafter, the DMHJ formula):
\begin{equation} \label{eq:demarco}
\sigma_\bs = \frac{L}{\sqrt{T}} \left( 1 + \frac{q}{L} + \frac{q^2+2}{2L^2} + \frac{q}{2L^3} + \cdots \right) \qtext{where} L=\sqrt{2|\log (K/x_0)|\,} \text{ and } q = N^{-1}(m_\sabr),
\end{equation} 
where $N^{-1}(\cdot)$ is the inverse  CDF of the standard normal distribution.

\citet{gulisashvili2018mass} express the mass at zero as a similar expectation over the distribution of $V$,
$$ m_\sabr(y_0, \vov) = \mathbb{E}\left(m_\cev\left(y_0\sqrt{V}\right)\right),
$$
where the mass at zero under the CEV model is given by
\begin{equation} \label{eq:cev_m0}
m_\cev(\sigma) = 
\bar{\Gamma}\left(\frac{x_0^{2\betac}}{2\betacpow{2}\sigma^2 T};\, \frac{1}{2\betac}\right) \qtext{for}
\bar{\Gamma}(x;a) = \bar{F}_\ncx(2x;2a,0) = \frac{1}{\Gamma(a)}\int_x^\infty t^{a-1} e^{-t} dt.
\end{equation}
The function $\bar{\Gamma}(\,\cdot\,;a)$ is the complementary CDF of the gamma distribution\footnote{This is equivalent to the upper incomplete gamma function normalized by the gamma function $\Gamma(a)$.} with the shape parameter $a$. In small-time limit, \citet{gulisashvili2018mass} cleverly approximate $V$ by a lognormal random variable through matching the first two moments, $\mu_1$ and $\mu_2$:
$$ V \approx \mu_1 \exp\left( \lambda Z - \frac{\lambda^2}{2} \right) \qtext{for} \lambda = \sqrt{\log\left(\frac{\mu_2}{\mu_1^2}\right)} \text{ and standard normal variate $Z$}.
$$
The first two moments of $V$ and, therefore, $\lambda$, are analytically available as
$$ \mu_1 = \frac{(w-1)}{\vov^2 T},\; \mu_2 = \frac{(w^6 - 6w + 5)}{15\vov^4 T^2}, \;\text{and}\;\lambda = \sqrt{\log\left(\frac{w^4 + 2w^3 + 3w^2 + 4w + 5}{15}\right)} \qtext{for} w=e^{\vov^2 T}.
$$
\citet{gulisashvili2018mass}, however, do not specify the method of numerical integration. 

Here, we employ the Gauss--Hermite quadrature (GHQ)~\citep[p.~890]{abramowitz} for more accurate integration. Let $\{z_k\}$ and $\{w_k\}$ for $k=1,\ldots,n$ be the points and weights, respectively, of the GHQ, which then can exactly evaluate the expectation of the polynomials up to the order of $2n-1$ with respect to the standard normal distribution. Using $\{z_k\}$ and $\{w_k\}$, the mass at zero is approximated as the weighted sum:
$$ m_\sabr(y_0, \vov) \approx \sum_{k=1}^n w_k\, m_\cev(y_0\sqrt{v_k}) \qtext{for}
v_k = \mu_1 \exp\left( \lambda z_k - \frac{\lambda^2}{2} \right).
$$

\begin{figure}[ht!]
	\caption{\label{fig:1} The mass at zero, $m_\sabr$, as a function of time to maturity, $T$. We test $(x_0, y_0, \vov)=(0.03, 0.6, 1)$, and $\beta=0.5$ (left) and $\beta=0.3$ (right) with $n=10$ Gauss--Hermite quadrature points.} 
	\vspace{2ex}
	\centering
	\includegraphics[width=0.48\textwidth]{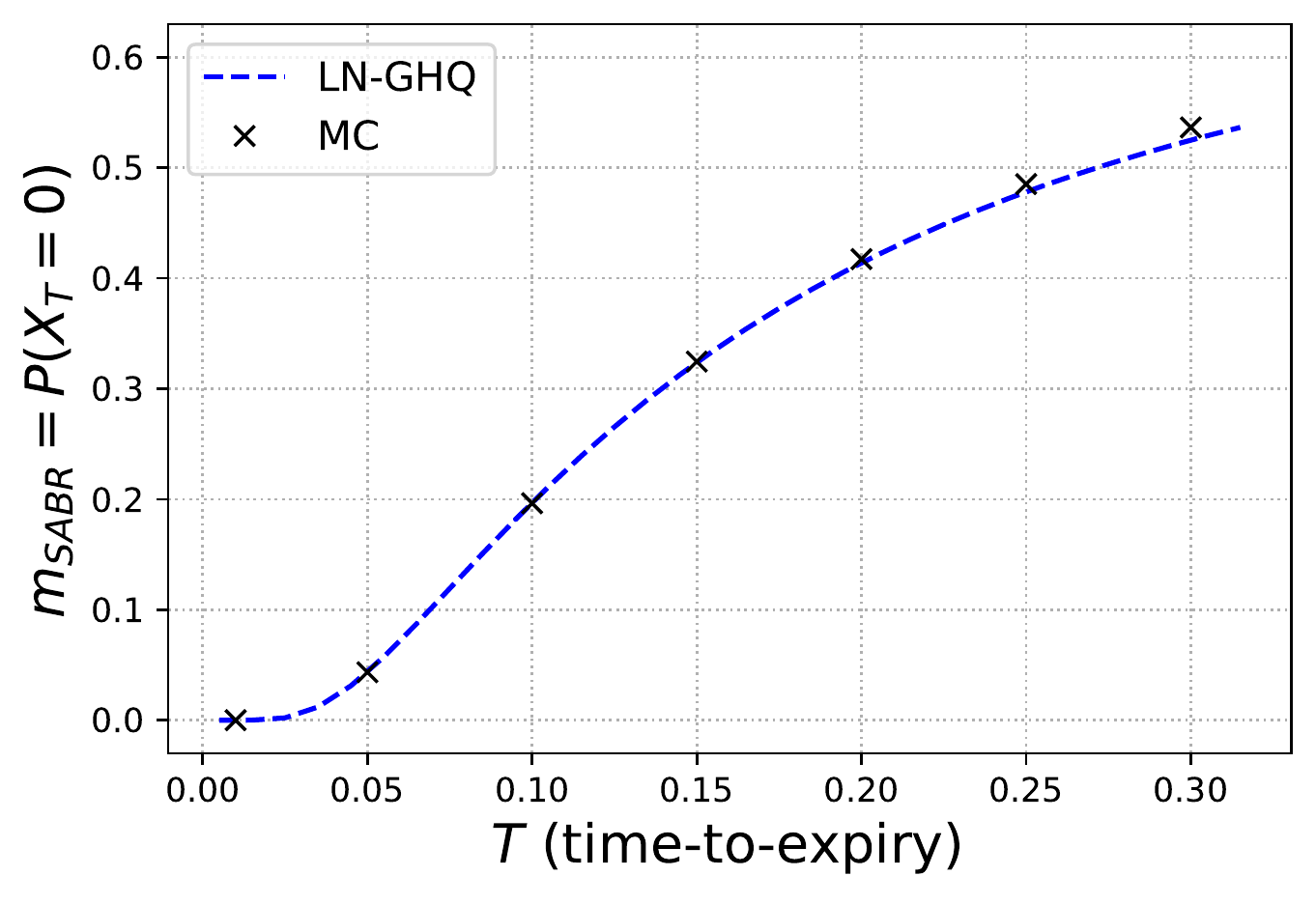}
	\includegraphics[width=0.48\textwidth]{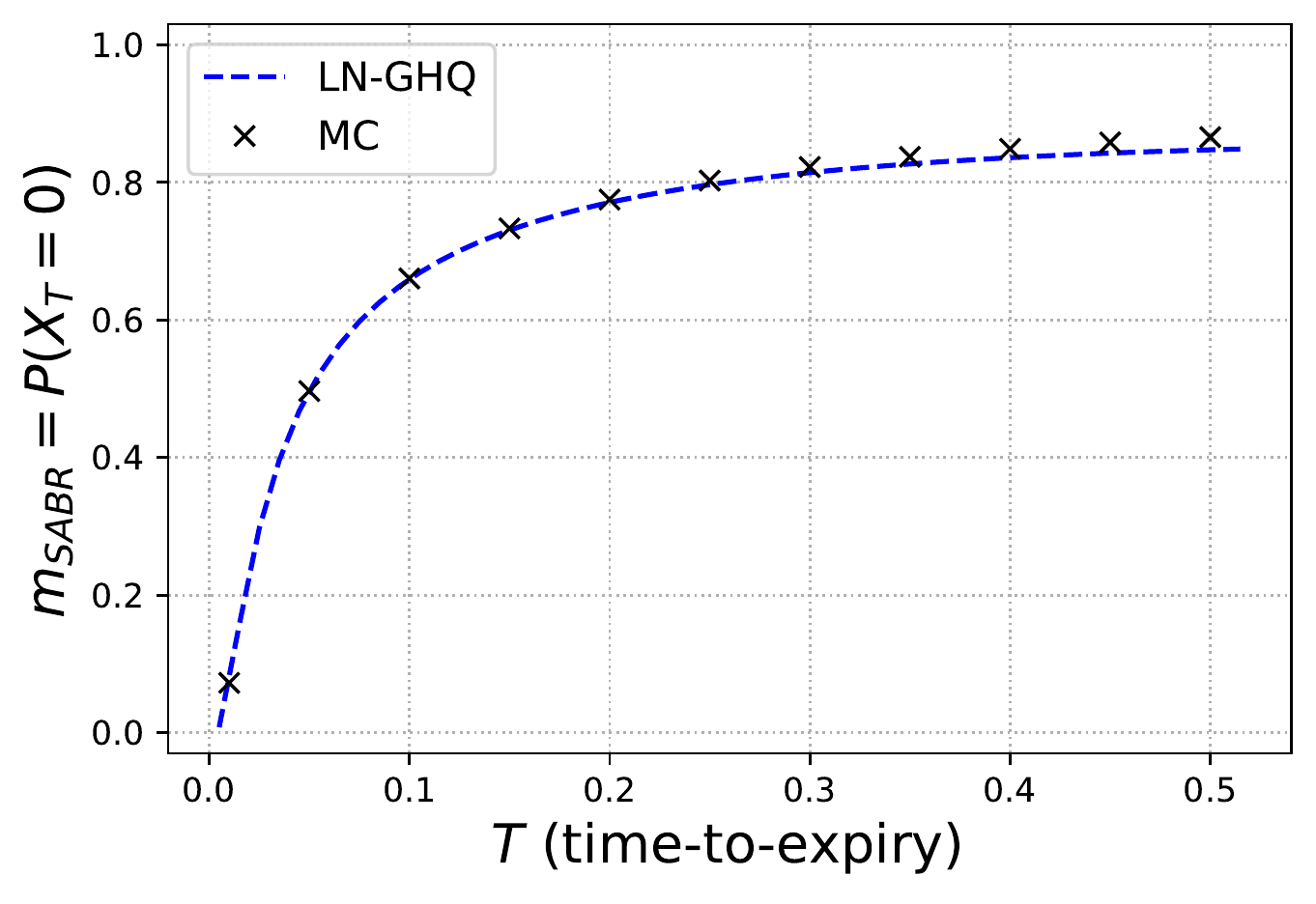}
\end{figure}

We demonstrate the performance of our method with numerical examples. For accuracy benchmarking, we implement the Monte-Carlo method as well. We draw random values of $V$ by simulating the paths of $Y_t$ over the discretized time steps and integrating the values with Simpson's rule. In Figure~\ref{fig:1}, we redo\footnote{Because the value of $x_0$ is not given in \citet{gulisashvili2018mass}, we guess $x_0=0.03$ based on the `Exact' label in the reference.} Figure~2 of \citet{gulisashvili2018mass}. Figure~\ref{fig:1} shows that the mass at zero computed with GHQ (labeled as LN-GHQ) is very close to the Monte-Carlo benchmark values (labeled as MC). This contrasts to Figure~2 of \citet{gulisashvili2018mass} where the `LN approx' label shows significant deviation from the `Exact' label. Our method also preserves the monotonicity of the mass at zero as $T$ increases. Consequently, we do not see the need for the `small-time' and `hybrid' approaches introduced in \citet{gulisashvili2018mass}. The convergence of the GHQ is known to be very fast. We use merely $n=10$ GHQ points for Figure~\ref{fig:1}. The errors from the converged values, evaluated with $n=100$, are in the orders of $10^{-9}$ and $10^{-7}$, respectively, for the two examples. See also \citet{choi2019sabrcev} for another numerical example and comparisons with other methods estimating the mass at zero.

\section{Option Price as an Integral of the CEV Option Price}
Moreover, we can similarly evaluate the option price under the SABR model. Using the GHQ, the price in Eq.~\eqref{eq:call_sabr} can be approximated by the weighted sum of the CEV option prices in Eq.~\eqref{eq:cev_call}:
\begin{equation} \label{eq:call_sabr_ghq}
C_\sabr (y_0, \vov) \approx \sum_{k=1}^n w_k C_\cev(y_0\sqrt{v_k}) \qtext{for} v_k = \mu_1 \exp\left( \lambda z_k - \frac{\lambda^2}{2} \right).
\end{equation}
This direct pricing approach has several critical advantages over the original method using the DMHJ formula. First, we can price the options for any strike price, while the DMHJ formula is accurate only for small strikes and diverges for bigger strikes. Second, the price from the new method is completely arbitrage-free because so is each of the CEV price components. Third, this approach can be applied to the lognormal ($\beta=1$) and normal ($\beta=0$) SABR models, to which the DMHJ formula is not applicable because the mass at zero is not available.\footnote{Under the lognormal SABR model, the origin is an unreachable boundary. Under the normal SABR model, the origin is not a boundary as the price can go negative. See \citet{antonov2015free} and \citet{choi2019nsvh} for more detail on the normal SABR model.} In those two cases, the CEV price in Eq.~\eqref{eq:call_sabr_ghq} should be replaced by the Black--Scholes and Bachelier prices, respectively.

\begin{figure}[ht!]
	\caption{\label{fig:2} The Black-Scholes volatility smile as a function of log strike price for two parameter sets: $(x_0, y_0, \vov, \beta, T)=(0.5, 0.5, 0.4, 0.5, 2)$ (left) and $(x_0, y_0, \vov, \beta, T)=(0.05, 0.4, 0.6, 0.3, 1)$ (right). The mass at zero, estimated by the method of this study, are $m_\sabr = 0.1657$ and $0.7624$, respectively.} 
	\vspace{2ex}
	\centering
	\includegraphics[width=0.48\textwidth]{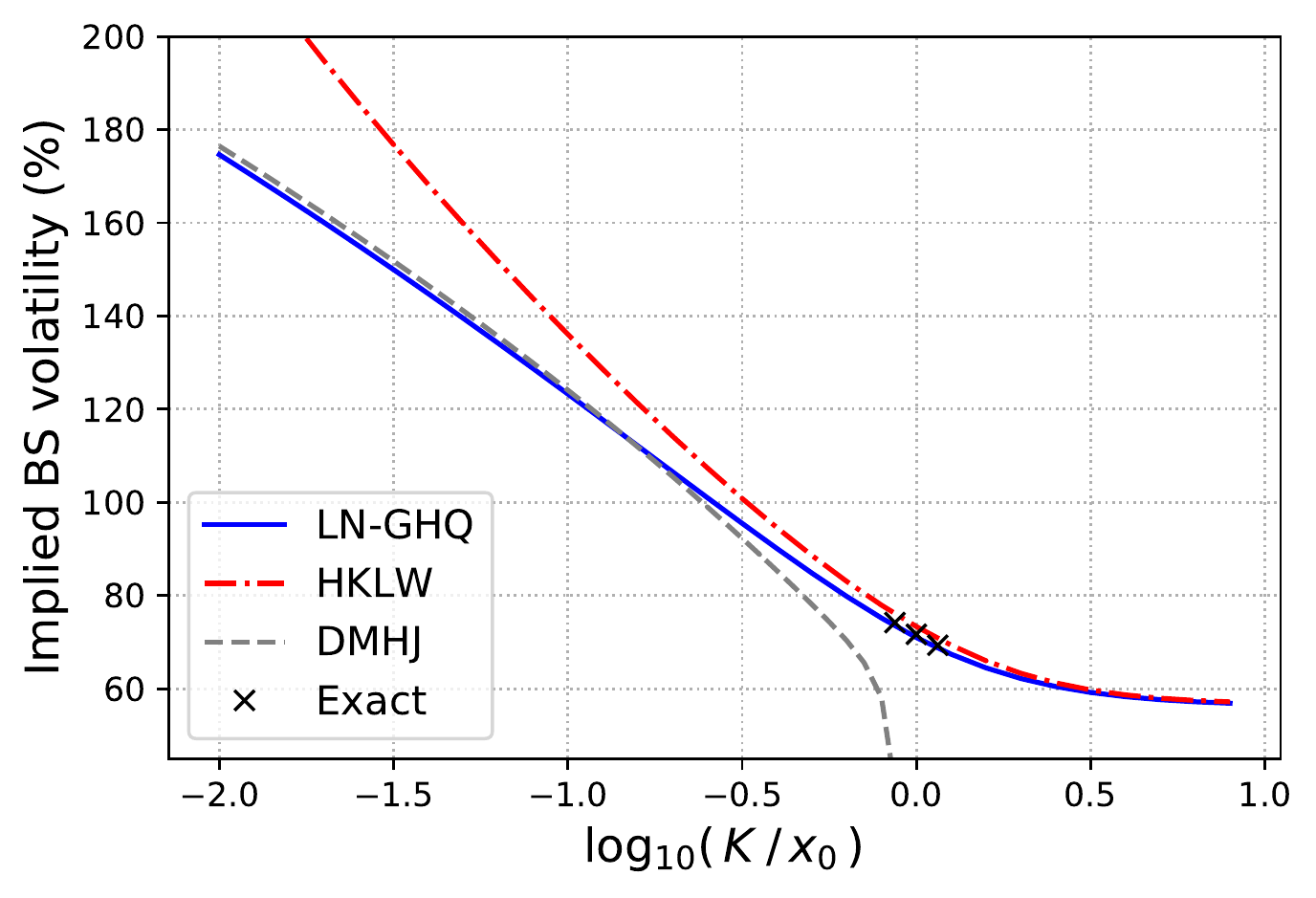}
	\includegraphics[width=0.48\textwidth]{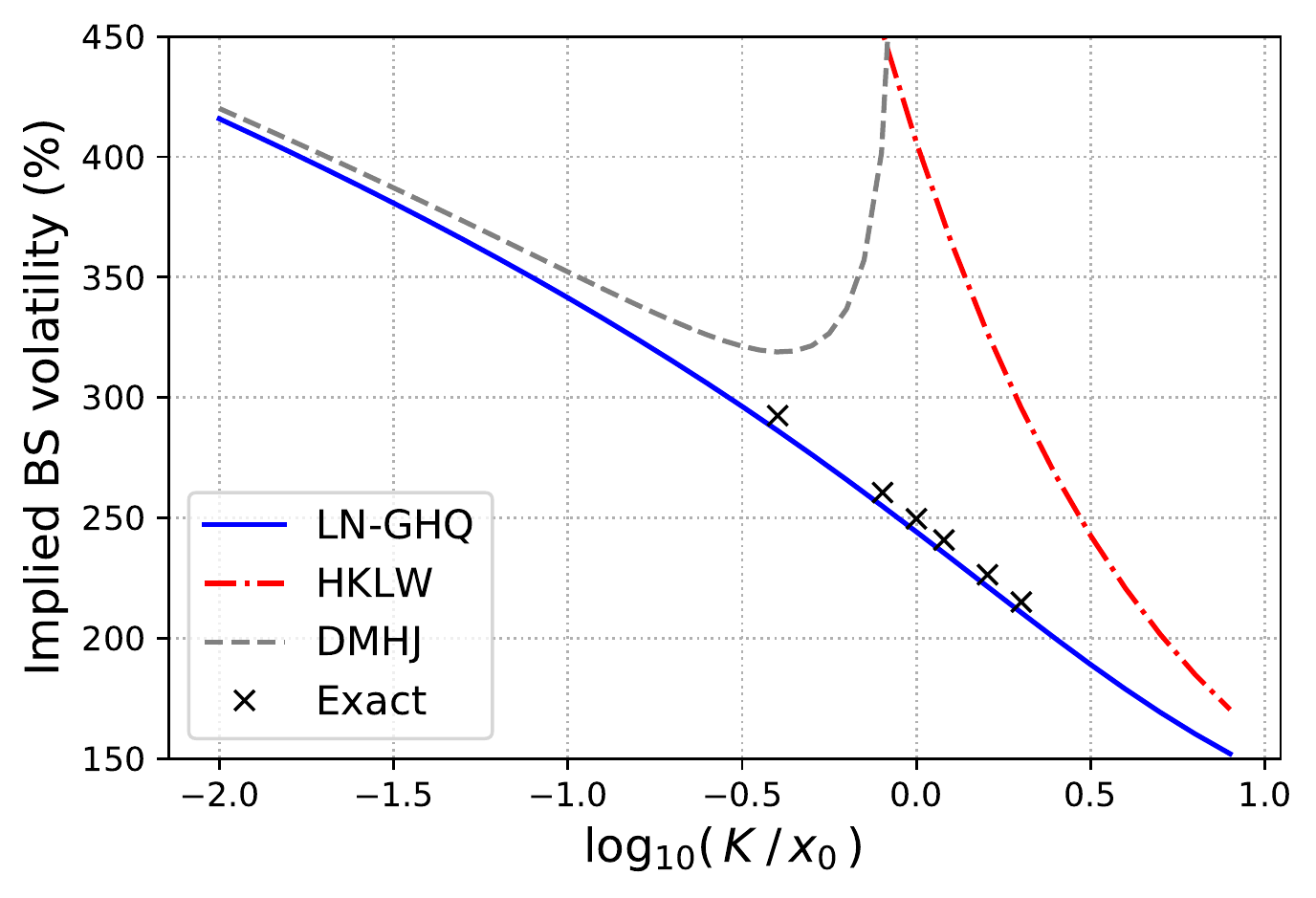}
\end{figure}

Figure~\ref{fig:2} demonstrates the advantages of the direct pricing method. We examine two parameter sets tested by prior studies, \citet{vonsydow2019benchop} and \citet{cai2017sabr}. The exact options prices are available from the references. We also compare the HKLW formula (labeled as HKLW) as it is the industry standard.\footnote{For other advanced volatility approximation methods, see \citet{choi2019sabrcev}.} The lognormal approximation with GHQ (labeled as LN-GHQ) shows an excellent agreement with the exact implied volatilities near the money. In the second parameter set (right), the HKLW formula significantly deviates from the exact value. In the low-strike region, the volatility smile from our method is also consistent with the DMHJ formula, with estimated mass at zero, $m_\sabr = 0.1657$ (left) and $0.7624$ (right), respectively. These are very close from the values from the Monte-Carlo method, $m_\sabr = 0.1634$ (left) and $0.7758$ (right) respectively. Yet, not surprisingly, the DMHJ formula diverges near the money ($K=x_0$). Therefore, our new pricing method is superior to the existing method across all strike range.

\section*{Funding} \noindent
Jaehyuk Choi gratefully acknowledges the financial support of the 2019 Bridge Trust Asset Management Research Fund. Lixin Wu was supported by Hong Kong RGC Grant \#16306717.

\newpage
\singlespacing
\bibliography{../../@Bib/SV_Z2}

\end{document}